\documentclass[twocolumn,trackchanges]{aastex701}

\usepackage{amssymb,amsmath}
\usepackage{threeparttable}
\usepackage{listings}
\usepackage{adjustbox}
\usepackage[T1]{fontenc}
\usepackage{soul}
\usepackage{xcolor}
\usepackage{pifont}

\begin{document}

\title{In Search of the Gravitational Wave of Fast X-Ray Transient in the Second Part of the LIGO-Virgo-KAGRA Fourth Observing Run}

\author[orcid=0000-0003-2658-5871]{Po-Ya Wang}
\affiliation{Institute of Astronomy, National Tsing Hua University, Hsinchu City, Taiwan, R.O.C}
\email[show]{poyawang@gapp.nthu.edu.tw}  

\author[orcid=0000-0002-5105-344X]{Albert K.H. Kong} 
\affiliation{Institute of Astronomy, National Tsing Hua University, Hsinchu City, Taiwan, R.O.C}
\affiliation{Institute of Space Engineering, National Tsing Hua University, Hsinchu City, Taiwan, R.O.C}

\email{}

\author[orcid=0000-0002-3333-8070]{Surojit Saha}
\affiliation{Institute of Astronomy, National Tsing Hua University, Hsinchu City, Taiwan, R.O.C}
\email{}

\begin{abstract}

Fast X-ray Transients (FXTs) are brief, intense X-ray flashes unassociated with persistent X-ray sources or known stellar objects, typically lasting from seconds to hours. With the launch of the Einstein Probe in January 2024, the study of FXTs has entered a new phase. Its large field of view significantly increases the discovery rate, enabling more robust statistics. The leading discoveries have revealed a fraction of gamma-ray bursts, while the rest remain unexplained. We propose a methodology to search for possible relations between gravitational waves (GWs) and FXTs, as it may serve as a discrepancy among proposed theoretical models. To achieve this, we intend to utilize X-Pipeline, an unmodeled search algorithm, to identify excess coherent signals in LIGO-Virgo-KAGRA (LVK) O4b observations. Out of 64 FXTs events, we obtained 19 FXTs for constructing robust searches. Given the absence of any detectable GW signals, we subsequently calculated upper limits and estimated exclusion distances for each source. In order to enrich the sample base and statistical analysis, this methodology could henceforth be conducted through future GW searches, including KAGRA in the O4c and O5 observations.

\end{abstract}


\section{Introduction} \label{sec:intro}

Fast X-Ray Transients\citep{2010csxs.book..267H} (hereafter FXTs) are characterized by abrupt X-ray emissions spanning timescales of seconds to several hours and are spatially distinct from any persistent X-ray sources or previously identified stellar objects. FXTs represent a heterogeneous class of short-lived high-energy events, whose physical origins remain under active investigation. Their observed diversity in peak luminosity, duration, spectral hardness, and host-galaxy environment suggests that multiple progenitor channels may contribute to the overall population. While the observed peak X-ray luminosity, $L_{\mathrm{X,Peak}}$, provides a useful first-order discriminator, it is not sufficient on its own to uniquely identify the underlying mechanism. Additional clues are encoded in the temporal evolution of the light curve, the thermal or nonthermal nature of the spectrum, the rise and decay timescales, and the localization of the event within or beyond its host galaxy. The major classes of proposed astrophysical origins are: stochastic outbursts from X-ray binaries (XRBs) \citep{1999ApJ...519...89C,2006ApJ...646..174K,miniutti2019nine}, shock breakout of core-collapse supernovae (CCSNe) \citep{2008Natur.453..469S,2020ApJ...896...39A,2022ApJ...927..224S,2023MNRAS.525.1562S}, off-axis gamma-ray bursts (GRBs) produced by a wide and mild jet \citep{2013ApJ...763L..22Z,2017MNRAS.467.4841B,2019Natur.568..198X,2021ApJ...915L..11A,2022ApJ...927..211L,2024arXiv240416350L,2024A&A...690A.101W}, tidal disruption events (TDEs) of systems with a white dwarf (WD) and an intermediate-mass black hole (IMBH) \citep{2013ApJ...779...14J,2014ApJ...794....9M,2015MNRAS.450.3765G,2019ApJ...884L..34P,2020SSRv..216...39M,2021SSRv..217...18S}, binary neutron star (BNS) mergers \citep{2017MNRAS.467.4841B,2019Natur.568..198X,2024A&A...683A.243Q}, proto-neutron stars in supernova models \citep{2009MNRAS.396.2038B,2019Natur.568..198X, 2023ApJ...948...91E,2003LRR.....6....2F}, collapsars in gravitationally unstable neutrino-cooled disks \citep{2024ApJ...972L...4G,2019MNRAS.487.4721Y,2023ApJ...948...91E,2006A&A...454...11R}, and cocoon breakout caused by jet-envelope collisions \citep{2018MNRAS.475.2971B,2017Sci...358.1559K,2018MNRAS.478..407N}.

It is therefore crucial to have statistically significant observations to support and help solve the degeneracy of these proposed scenarios. Previously, XMM-Newton \citep{2001A&A...365L...1J} and Chandra \citep{2002PASP..114....1W} were the instruments that discovered the majority of FXTs, resulting in 12 and 22 events, respectively. Following the successful deployment of the \textit{Einstein Probe} (EP) in early 2024, equipped with a $3600$ square-degree field-of-view soft (0.5--4 keV) X-ray telescope \citep{2025SCPMA..6839501Y}, the observed number of FXTs is now 4 times greater than in the previous 20 years. With higher sensitivity in the soft X-ray band and improved temporal resolution, it is expected to uncover the intrinsic luminosity function and event-rate distribution of FXTs, enabling a physically grounded taxonomy rather than a purely phenomenological one. Interestingly, several FXTs are spatially and temporally coincident with GRBs \citep{2025ApJ...986..106G}.

The first triggered FXT detection, EP240219a\citep{2024ATel16463....1Z}, was indeed followed by GRB 240219A\citep{2024ApJ...975L..27Y}. Based on the spectral analysis of EP and the Fermi Gamma-ray Space Telescope, it is shown that EP240219a/GRB240219A is an X-ray bright GRB. EP240315a is another notable event, being located in a z$\sim4.9$ host galaxy\citep{2025NatAs...9..564L,2024arXiv240416350L,2024ApJ...969L..14G}. EP observed a significant signal with a peak luminosity of $\sim10^{51}$erg\ s$^{-1}$. Together with follow-up {\it Swift} and {\it KONUS-Wind} observations, EP240315a is confirmed to be a GRB. Optical spectroscopic observations by the Very Large Telescope/X-shooter and Gemini North/GMOS-N provided precise estimates of the redshift and served as a rare case of clear localization of a distant FXT/GRB.

It is widely accepted that supernovae (SNe) serve as progenitors for a distinct class of GRBs. In this context, we present a brief overview of notable FXTs with established SN associations: EP250108a is a counterpart of SN 2025kg, a broad-line Type Ic SN located at z$\sim0.18$. Although a GRB did not accompany this source, its optical counterpart, observed by JWST/NIRSpec shows similarities to GRB-SNe \citep{2025arXiv250408889R}. Earlier in 2024, EP240414a was also shown to be a broad-line Type Ic SN through its extremely blue optical spectra\citep{2025ApJ...982L..47V}.

Given the established connection between FXTs and GRBs, and their association with compact objects such as SN, we were motivated to investigate whether gravitational wave (GW) counterparts might exist. Following the successful multi-messenger observation of the BNS merger GW170817 \citep{2017PhRvL.119p1101A}, the search for GRBs associated with gravitational-wave (GW) events has emerged as a key objective in multi-messenger astronomy. The LIGO-Virgo-KAGRA (LVK) collaboration utilizes two main pipelines, PyGRB\citep{2011PhRvD..83h4002H, 2014PhRvD..90l2004W} and X-pipeline\citep{2010NJPh...12e3034S,2012PhRvD..86b2003W}, to analyze offline GW strain data, specifically searching for gravitational wave signals that are temporally and spatially coincident with observed GRBs. PyGRB is a targeted, matched-filtering analysis designed to uncover modeled signals from compact binary coalescences. Conversely, X-pipeline is an unmodeled burst search algorithm that looks for coherent excess power across the detector network.

 The main difference between the two searches is the assumption about the GW signal morphology. In principle, PyGRB requires an assumption of waveform templates, i.e., deciding whether it is a BNS, binary black hole, or neutron star-black hole system. Although the template bank serves a large parameter space, it would still limit the search to certain physical origins. Due to the uncertainty regarding the physical nature of FXTs, PyGRB may not be a suitable choice. In contrast, X-pipeline does not assume signal morphology when searching through combined networks of GW detectors. It coherently searches for excess signals from the datasets according to the time and spatial coordinates of GRB events. It is therefore a better choice for the search for FXT events. In this work, we employed X-pipeline to search for selected FXTs observed by the Einstein Probe during the second part of the 4th observing run (from 2024-04-10 to 2025-01-28) of Advanced LIGO and Advanced Virgo (O4b).

This paper is organized as follows. Section \ref{sec:method} briefly describes the search algorithm and parameter settings of X-pipeline. Section \ref{sec:sample} discusses how the FXT samples were selected and summarizes their basic properties. In Section \ref{sec:result}, we present the results obtained from the X-pipeline with the FXT samples and provide a discussion combining the GW search and multi-wavelength observations. Finally, Section \ref{sec:conclu} provides the concluding remarks, including comments on the significance of the results and proposed future follow-ups.

\section{Methodology} \label{sec:method}

In this section, we outline the GW signal search algorithm of the X-pipeline \citep{2010NJPh...12e3034S} and introduce the parameter settings for our search of FXTs.

\subsection{General Setup and Parameter Settings}
Here we present the general setup of X-pipeline. Serving as a targeted search algorithm, several user-defined preliminaries are required: a. The set of detectors, b. the range of the on-source window (the possible time window for a GW occurrence), c. data streams (channels), d. the set of sky positions (the FXT position), and e. the list of numerical method parameters (e.g., Fast Fourier Transform length, size of analyzing block, etc.). We explain the details as follows.

The set of detectors used in this work includes all three operating detectors in the O4b run: LIGO-Hanford (H1), LIGO-Livingston (L1), and Virgo (V1). An FXT event will be analyzed if the burst occurred at all three detectors or at both H1 and L1, which are operating and have a sufficiently long data stream.

The definition of a sufficiently long data stream relates to the definitions of the on-source window and off-source window here. In this analysis, the on-source window refers to the expected arrival time of the GW counterpart relative to the targeted electromagnetic burst. In contrast, the off-source window is the data stream during which we expect no GW events to occur, referred to as the background. A 10,800-second off-source window is set, centered on the triggered time of each event\citep{2010NJPh...12e3034S}. In this work, we apply an on-source window as [T$_0-600$s, T$_0+60$s], where T$_0$ refers to the event time obtained from the Einstein Probe, which is the same window in the previous GRB-GW counterpart searches \citep{2021ApJ...915...86A, 2022ApJ...928..186A}. This is based on the fact that, according to current observations, one-third of FXTs are spatially and temporally associated with GRB. The samples of this paper will be described in section \ref{sec:sample}. Consequently, this study searches for GW counterparts by operating under the premise that FXTs and GRBs are driven by fundamentally similar astrophysical mechanisms.

The set of sky positions of FXTs comes with the observation processed by the Einstein Probe. The positional errors are also considered, which are around $0.05$ degrees. Readers can refer to Section \ref{sec:sample} for further individual observation details. Finally, the numerical parameter settings are established as follows. The lengths of the fast Fourier transforms are defined as $2.0, 1.0, 0.5, 0.25, 0.125, 0.0625, 0.03125, 0.015625,$ and $0.0078125$ seconds. Each transform block is determined to be $256$ seconds, and each of the analysis blocks covers half of the preceding ones. The search frequency range could also be defined; here, we set it to [$20$, $500$] Hz, targeting the most sensitive range of our detectors. The main operation of the X-pipeline starts with the construction of the time-frequency map for each block, aiming to produce a coherent set of $h_\times$ and $h_+$ according to the maximum likelihood method. According to Figure 2 in \cite{2010NJPh...12e3034S}, the intensity of the highest one percent of pixels would then be clustered and defined as a candidate event. Both on-source and off-source windows will be used to perform a search for candidate events. A list of events from both windows will be output along with the significance according to the $\chi^2$ distribution.

Up to now, all the on-source candidate events should undergo a glitch rejection process. While the null incoherent energy $I_{\text{null}}$ should be similar to the null coherent energy $E_{\text{null}}$ from the on-source candidate events, they should also be distinguishable from the events identified in the off-source window. A measure of how far an on-source candidate event is from the median of the off-source window events can be used as a measure of determination. Here, a function is defined to serve as a threshold:
\begin{align}\label{eq1}
\log_{10}I=a(\log_{10}E)^2+c\tt{.}
\end{align}

The parameters of $a$ and $c$ in the curve of the cut are fitted to the median values of $\log_{10}I$ and $\log_{10}E$, where I and E represent the coherent energy and incoherent energy, respectively. While this is an artificially selected curve, it distinguishes the background (off-source) events well in practice. An example plot could be found in Figure 4 of \cite{2010NJPh...12e3034S}.

Once all the glitch rejection processes for the on-source event are performed, the highest significance source remaining is called the loudest event. To claim that the loudest event is truly a GW detection, we further need to be confident that it is statistically inconsistent with the background. Here, for each off-source data segment, we also obtain the loudest event, which would form a set of significant background events. A p-value of the on-source loudest event compared to the background loudest events could then be calculated, giving the probability of such an event being 'not' a GW detection, i.e., the probability under a null hypothesis. The derived values and discussion will be presented in Section \ref{sec:result}.

\subsection{Upper Limit Estimation and Exclusion Distance}\label{sec:para}

Since there is a high chance that the p-value of the loudest event does not imply detection of GW, an upper limit calculation is performed after estimating the probability.

In this work, we select simulated waveforms that are physically motivated to produce both electromagnetic and GW emissions to inject into the off-source window:
\begin{enumerate}
    \item Circular Sine–Gaussian (CSG): bursts intended to represent GW emission from stellar collapse, defined as in \citet{2017ApJ...841...89A}. We use the quality factor, defining the number of oscillations before a burst fades, as $Q = 9$ and central frequencies of 70, 100, 150, and 300 Hz. In all cases, an optimistic GW energy emission of $E_{\mathrm{GW,90\%}} = 10^{-2} M_{\odot}c^{2}$ is assumed.
    \item Compact binary inspirals: binary neutron star (BNS) and neutron star–black hole (NSBH) mergers, modeled with Gaussian mass distributions centered at $1.4\,M_{\odot}$ for NSs (width $0.2\,M_{\odot}$ for BNS, $0.4\,M_{\odot}$ for NSBH). These represent plausible progenitors of short GRBs \citep{2017ApJ...841...89A}.
    \item Accretion disk instability (ADI): long-duration signals associated with instabilities in a magnetically suspended torus around a rapidly spinning black hole \citep{2001PhRvL..87i1101V, 2014MNRAS.444L..58V}. We adopt the five ADI waveform families used in \citet{2017ApJ...841...89A}.
\end{enumerate}

We quantify the sensitivity of the generic transient search by injecting simulated signals into the off-source data and attempting to recover them. For each waveform family, we determine the percentage of injections that yield a significance higher than the loudest on-source event. The amplitude at which 90\% of the injections exceed this threshold is taken as the upper limit. Calibration uncertainties are included by jittering the amplitude and arrival time of injections according to a Gaussian distribution representative of LVK calibration errors \citep{2017ApJ...841...89A, 2021arXiv210700129S, 2022CQGra..39d5006A}. These injection studies can further allow us to calculate the 90\% exclusion distances for each possible waveform, which will help interpret the nature of FXTs. A combined result with multi-wavelength observations will be discussed in Section \ref{sec:result}.

\section{Samples} \label{sec:sample}

This section details the Fast X-ray Transient (FXT) sample and the selection criteria employed in this work. First, we compiled a list of FXTs observed by the Einstein Probe, sourced from the General Coordinates Network (GCN) circulars\footnote{https://gcn.nasa.gov/circulars/}. Because our objective is to identify gravitational-wave (GW) counterparts detected by the LIGO-Virgo-KAGRA (LVK) network, we restricted our selection to FXT events occurring during the O4b observing run (specifically, from 2024-04-10 15:00 UTC to 2025-01-28 17:00 UTC). This initial search yielded 64 FXT events, for which we recorded the identifier, trigger time (in UTC), sky coordinates (Right Ascension and Declination), and positional uncertainty in degrees. For our analysis, we required that the LVK network have either all three detectors online or, at a minimum, both the Hanford (H1) and Livingston (L1) interferometers available at the time of the event. Furthermore, a subset of sources associated with GRBs has been reserved for an independent study (LVK Collaboration, in prep.) and is therefore excluded from the current search. Applying these criteria to the initial $64$ sources, we identified 11 events with all three detectors available and an additional $8$ events with only H1 and L1 available. The final source list selected for the X-pipeline analysis is summarized in Table~\ref{tab:final}. For reference, the complete list of these 64 events is provided in Section \ref{A:sample list}.

\begin{table*}[htp]
    \centering
    \caption{Final FXT list for analysis sorted in ascending order based on detection time. The sources for the three-detector analysis are marked with an uppercase star, while the others are used for the two-detector analysis.}
    \label{tab:final}
    \begin{threeparttable}
        \begin{tabular}{lccccc}
        \toprule
            Source Name & Triggered time & R.A. & Dec & Positional Error\\
            & (UTC time) & (deg) & (deg) & (deg) & \\
            \hline
            EP240413a \tnote{*} & 2024-04-13T14:39:37 & 228.794 & -18.800 & 0.333\\
            EP240414a \tnote{*} & 2024-04-14T09:50:12 & 191.498 & -9.695 & 0.050\\
            EP240426a \tnote{*} & 2024-04-26T04:23:21 & 121.857 & -29.461 & 0.003\\
            EP240426b & 2024-04-26T14:19:06 & 173.787 & -40.741 & 0.050\\
            EP240506a \tnote{*} & 2024-05-06T05:01:39 & 213.978 & -16.715 & 0.050\\
%            EP240518a & 2014-05-14T13:02:08 & 216.955 & -49.565 & 0.050\\
            EPJ0052.9-7230\tnote{*} &2024-05-27T08:41:28&13.215& -72.494&0.050\\
            EP240617a \tnote{\textdagger,*} & 2024-06-17T12:19:13 & 285.030 & -22.561 & 0.050\\
            EP240625a \tnote{*} & 2024-06-25T01:48:23 & 310.760 & -15.966 & 0.033\\
            EP240626a \tnote{*} & 2024-06-26T06:28:28 & 263.023 & -13.051 & 0.033\\
            EP240708a \tnote{*} & 2024-07-08T23:28:23 & 345.963 & -22.840 & 0.050\\
            % EP240913a \tnote{\textdagger} & 2024-09-13T11:39:33 & 16.681 & 16.75 & 0.042\\
            EP240918c & 2024-09-18T18:06:47 & 281.338 & -13.167 & 0.039\\
            EP240919a \tnote{\textdagger} & 2024-09-19T14:47:40 & 334.2797
 & -9.7362 & 0.006\\
            % EP241109a & 2024-11-09T06:01:55 & 18.3599 & 0.0184 & 0.004\\
            EP241125a & 2024-11-25T00:06:06 & 48.561 & 37.677 & 0.044\\
            EP241201a & 2024-12-01T21:01:22 & 282.596 & 66.081 & 0.039\\   
            EP241212a\tnote{*} &2024-12-12T11:21:52&153.817&60.068&0.042\\
            % EP241213a \tnote{\textdagger,*} & 2024-12-13T02:17:15 & 116.182 & 35.271 & 0.047\\
            EP241231b & 2024-12-31T19:27:29 & 100.064 & 16.171 & 0.043\\
            EP250101a\tnote{*} &2025-01-01T05:52:21&85.575&0.352&0.053\\
            EP250108a & 2025-01-08T12:30:28 & 55.623 & -22.509 & 0.037\\
            EP250125a & 2025-01-25T02:37:31 & 175.364 & -21.708 & 0.038\\
            \hline
        \end{tabular}
        \begin{tablenotes}
            \item[\textdagger] Identified as a GRB.
            \item[*] Observed by H1, L1, and V1.
        \end{tablenotes}
    \end{threeparttable}
\end{table*}

\section{Results and Discussion} \label{sec:result}

With the final source list described in Section~\ref{sec:sample}, detailed in Table~\ref{tab:final}, we now present the results from the X-pipeline. The primary outcome from the pipeline provides a statistical probability estimation, or p-value, regarding the "non-existence" of a GW counterpart for each event when compared to the background. For each source, the analysis yields (i) a sensitivity upper limit based on a null hypothesis and (ii) a distance lower bound, referred to as the exclusion distance, derived for potential models. In this section, we will first individually address the details of notable sources and then discuss the statistical behavior.

\subsection{Individual Observations}
In this subsection, we discuss the results from X-pipeline searches and the archived follow-up observations of sources associated with notable transient events, such as supernovae or subthreshold GRBs. For all the FXTs discussed here, we assumed that they were initially detected or triggered by the Wide-Field X-ray Telescope onboard the EP, unless stated otherwise (see also Section~\ref{A:sample list} Table~\ref{tab:grb_all}). All of the optical and infrared magnitudes are in AB magnitudes, and the time is in UTC. For the sources with redshift measurements, we provide the original report for reference in the following. A general discussion of EP sources' distance may refer to \cite{2025ApJ...993L..37O}.

\begin{table*}[htp]
    \centering
    \caption{The Exclusion Distances of FXT sources.}
    \label{tab:exclusion distances}
    \begin{tabular}{l|ccccccccccc}
        &\multicolumn{11}{c}{Max Exclusion (Mpc) by Injection Model}\\
        \hline
        HLV Sources & sgc70Q9 & sgc100Q9 & sgc150Q9 & sgc300Q9 & bns & nsbh & adi\_a & adi\_b & adi\_c & adi\_d & adi\_e \\
        \hline
        EP240413a & 249.7 & 226.3 & 158.0 & 79.8 & 68.9 & 152.5 & 51.7 & 204.2 & 76.4 & 32.5 & 78.1 \\
        EP240414a & 248.9 & 239.1 & 164.6 & 85.8 & 69.9 & 153.1 & 55.5 & 211.1 & 82.1 & 33.2 & 79.8 \\
        EP240426a & 232.0 & 176.8 & 126.2 & 63.6 & 52.8 & 137.0 & 46.5 & 156.7 & 68.3 & 23.9 & 55.0 \\
        EP240506a & 301.1 & 249.1 & 180.7 & 97.6 & 76.5 & 190.9 & 70.6 & 237.7 & 102.5 & 34.4 & 89.0 \\
        EPJ0052.9-7230 & 317.9 & 257.5 & 207.6 & 115.7 & 83.1 & 204.1 & 74.8 & 261.7 & 107.3 & 41.1 & 107.4 \\
        EP240617a & 247.4 & 235.6 & 163.4 & 82.0 & 70.3 & 150.5 & 49.7 & 204.5 & 75.3 & 31.8 & 75.1 \\
        EP240625a & 243.8 & 205.7 & 165.3 & 84.9 & 72.0 & 161.5 & 56.0 & 208.7 & 77.0 & 32.4 & 78.3 \\
        EP240626a & 284.6 & 245.7 & 184.4 & 109.6 & 75.1 & 201.6 & 69.9 & 226.9 & 101.2 & 34.5 & 83.1 \\
        EP240708a & 398.4 & 356.3 & 254.3 & 120.6 & 108.8 & 258.8 & 100.8 & 322.3 & 147.3 & 50.1 & 120.0 \\
        EP241212a & 338.2 & 266.6 & 241.9 & 122.6 & 104.2 & 235.5 & 100.3 &  310.5 & 137.1 & 48.7 & 122.0 \\
        EP250101a & 359.8 & 335.7 & 250.1 & 130.3 & 104.3 & 233.0 & 96.2 & 303.9 & 117.1 & 47.1 & 115.3 \\
        \hline
        HL Sources & & & & & & & & & & & \\
        \hline
        EP240426b & 406.9 & 354.4 & 263.4 & 141.9 & 111.8 & 269.7 & 103.5 & 330.6 & 149.0 & 50.3 & 126.3 \\
        EP240918c & 235.8 & 190.8 & 162.3 & 84.8 & 52.8 & 152.5 & 56.9 & 205.6 & 75.1 & 31.4 & 78.8 \\
        EP240919a & 340.8 & 267.8 & 232.3 & 121.0 & 69.0 & 213.7 & 77.0 & 294.0 & 108.5 & 39.2 & 107.2 \\
        EP241125a & 246.1 & 230.7 & 166.8 & 90.6 & 71.5 & 167.6 & 62.4 & 205.4 & 79.7 & 33.0 & 81.0 \\
        EP241201a & 360.7 & 312.7 & 234.3 & 116.2 & 103.1 & 234.2 & 88.4 & 300.9 & 110.1 & 47.0 & 115.0 \\
        EP241231b & 264.8 & 241.8 & 173.7 & 97.1 & 74.2 & 168.1 & 69.8 & 211.8 & 87.3 & 32.2 & 80.4 \\
        EP250108a & 337.7 & 255.1 & 230.3 & 117.9 & 100.8 & 217.7 & 70.1 & 273.9 & 97.4 & 37.0 & 101.9 \\
        EP250125a & 243.9 & 211.4 & 163.4 & 87.7 & 71.8 & 160.5 & 68.4 & 210.5 & 80.5 & 32.5 & 82.4\\
    \end{tabular}
\end{table*}

\subsubsection{EP240414a}\label{EP240414a}

EP240414a is a notable X-ray source that has garnered considerable interest due to its association with the optical transient AT2024gsa. It was first observed by EP at R.A., Dec.$= 191.498,-9.695$ deg, with a positional error of $3$ arcmin on 2024-04-14T09:50:12 \citep{2024GCN.36091....1L}. The EP Follow-up X-Ray Telescope observed the source 2 hours later and found that the flux had faded by 4 orders of magnitude \citep{2024GCN.36129....1G}. $3.13$ hours later, the Lulin One-meter Telescope identified the optical transient $1.56$ arcmin away from the FXT. This source is later confirmed to be located in the host galaxy SDSS J124601.99-094309.3, at z=$0.41$ \citep{2024GCN.36094....1A,2024GCN.36355....1L}. By Pan-STARRS and other observations days after the EP detection, a rise in multiple optical bands indicated the presence of a type Ic-BL supernova \citep{2025NatAs...9.1073S,2025ApJ...982L..47V,2025ApJ...978L..21S}. 

Assuming a cosmology parameter of $\text{H}_0 = (67.4 \pm 0.5) \text{km s}^{-1} \text{Mpc}^{-1}$ and $\Omega_m=0.315 \pm 0.007$ \citep{2020A&A...641A...6P}, the corresponding luminosity distance of z$\sim0.41$ is $2312.6$ Mpc, which is much larger than the upper limit we measured in Table \ref{tab:exclusion distances}. Therefore, the null detection of GW from this source is in agreement with the estimated p-value of $0.068$.

Furthermore, we can extend this discussion to constrain the gravitational-wave emission efficiency. For the CSG waveforms, the exclusion distances reported in Table\,\ref{tab:exclusion distances} assume a standard emission energy of $E_{\mathrm{GW}} = 10^{-2} M_{\odot}c^{2}$. However, because the true distance to this source is known, we can invert this relationship using Equation\,\ref{eq2} \citep{2013arXiv1304.0210S} to establish a 90\% upper limit on emission efficiency ($E_{\mathrm{GW,90\%}}$). The resulting limits, together with the corresponding $h_{\mathrm{rss,90\%}}$ values, are summarized in Table\,\ref{tab:energy limits-240414}. For CBC waveforms, we cannot perform a similar efficiency conversion because the strain output is dependent on the binary masses in general relativity; the corresponding $h_{\mathrm{rss,90\%}}$ limits for BNS and NSBH are also listed in Table\,\ref{tab:energy limits-240414}. Similarly, all ADI waveforms in this study were generated at a fixed $E_{\mathrm{GW}}$ governed by their intrinsic physical properties, radiating $0.02 M_{\odot}c^{2}$ for \texttt{adi\_a}, $0.22 M_{\odot}c^{2}$ for \texttt{adi\_b}, $0.25 M_{\odot}c^{2}$ for \texttt{adi\_c}, $0.02 M_{\odot}c^{2}$ for \texttt{adi\_d}, and $0.17 M_{\odot}c^{2}$ for \texttt{adi\_e}.

\begin{table}
    \centering
    \caption{The 90\% upper limits on $E_{\mathrm{GW,90\%}}$ and $h_{\mathrm{rss,90\%}}$ for the CSG and CBC waveforms for the search of EP240414a.}
    \label{tab:energy limits-240414}
    \begin{adjustbox}{width=\columnwidth,center}
    \begin{tabular}{lccc}
        \hline
        Waveform & $f_{\mathrm{central}}$ (Hz) & $E_{\mathrm{GW,90\%}}$ ($M_{\odot}c^{2}$) & $h_{\mathrm{rss,90\%}}$ (Hz$^{-1/2}$) \\
        \hline
        \multicolumn{4}{l}{\textbf{CSG}} \\
        CSG & 70  & $8.088\times10^{-1}$ & $6.205\times10^{-23}$ \\
        CSG & 100 & $8.754\times10^{-1}$ & $4.519\times10^{-23}$ \\
        CSG & 150 & $1.845$              & $4.374\times10^{-23}$ \\
        CSG & 300 & $6.818$              & $4.204\times10^{-23}$ \\
        \hline
        \multicolumn{4}{l}{\textbf{CBC}} \\
        BNS  & -- & -- & $1.645\times10^{-22}$ \\
        NSBH & -- & -- & $3.114\times10^{-22}$ \\
        \hline
    \end{tabular}
    \end{adjustbox}
\end{table}

\begin{equation}\label{eq2}
    E_{\mathrm{GW},90\%} = \frac{2}{5}\frac{\pi^2c^3}{G}r^2f_0^2h_{rss,90\%}^2
\end{equation}

\subsubsection{EP240426a}

EP240426a (R.A., Dec=121.8567,-29.4609 deg; 2024-04-26T04:23:21) was first reported by the EP Follow-up X-Ray Telescope \citep{2024GCN.36313....1S} as a candidate counterpart of the GW event S240422ed \citep{2024GCN.36236....1L}, a compact binary merger event on 2024-04-22T21:35:13.417. Although the FXT arrived $4$ days later, the sky position was located within $90\%$ of the GW localization error. A galaxy, 2MASX J08072584-2927344, 5.5 arcsec away from the FXT position, was measured to have a luminosity distance of $188$ Mpc, which was comparable to the GW event and considered to be a candidate host galaxy.

The EP team estimated an unabsorbed flux of $9.2 \times 10 ^{-13}$ erg/s/cm$^2$ ($0.5-10$ keV). About $8$ hours after the EP event, Swift-XRT also spotted a burst $6.2$ arcsec away with a flux of $7.0 \times 10^{-13}$ erg/s/cm$^2$. GW-MMADS DECam follow-up detected an optical transient, namely AT2024hhs, a day before the EP event, with a DECam-z filter magnitude of $21.01\pm0.10$. However, this variability might be linked to the AGN activity of the host galaxy, which may not be significantly related to the X-ray event.

As the parameter of the X-pipeline we have settled, based on the EP event on April $26$, the $+60\sim-600$ seconds on-source window does not occur for the GW event S240422ed \citep{2024GCN.36236....1L}, resulting in a reasonable non-detection, followed by a p-value of $0.662$. The exclusion distances of this source, shown in Table\,\ref{tab:exclusion distances} may not be accurate.
We attempted to produce a search using the triggered time of S240422ed; however, it was unfortunate that the surrounding background (off-source) time segments were not sufficiently long to complete a successful run. 

\subsubsection{EP240617a}

EP detected EP240617a at R.A., Dec.$=285.030, -22.561$ deg on 2024-06-17T12:19:13. The duration had a lower limit of $300$ seconds due to Earth occultation. The estimated unabsorbed flux was $3.5(+/-0.3) \times 10^{-9}$ in the $0.5-4$ keV band, with a peak flux of $1.4(+/-0.1) \times 10^{-8}$ erg/s/cm$^{2}$ \citep{2024GCN.36691....1Z}. The EP team claimed that there was an untriggered, subthreshold GRB detection in the Fermi/GBM data that overlapped with the time interval of EP240617a and was located within the error circle \citep{2024GCN.36692....1Y}. A software tool, RapidGBM, was then developed to visualize the GBM data and perform further analysis. As this tool was utilized for EP240617a, the authors suggest this event as an X-ray-rich gamma-ray burst \citep{2025ApJ...993...51W}.

Fortunately, there was a Swift-XRT detection $77$ hours after the EP trigger. They observed a weak signal at 5.3 sigma significance within the error radius of the EP detection. As the flux is about $1.9 \times 10^{-13}$ ergs/s/cm$^{2}$, this Swift-XRT detection is considered as a candidate X-ray afterglow following the FXT event \citep{2024GCN.36722....1S}.

For our X-pipeline results, we obtained a p-value of $0.593$ and exclusion distances shown in Table \ref{tab:exclusion distances}, indicating a non-detection of GW.

\subsubsection{EP240919a}

EP240919a is the second source, which was indicated as a GRB event. A detection by the Wide Field X-Ray Telescope was triggered at 2024-09-19T14:47:40, and the latter-triggered EP Follow-up Telescope located the source at R.A., Dec.$=334.2797, -9.7362$ deg with $20$ arcsec uncertainty. The duration exceeded $400$ seconds, with an absorbed flux of $7.0(+1.8, -1.5)\times 10^{-10}$ erg/s/cm$^2$ \citep{2024GCN.37561....1L}. Eighty seconds after the trigger time by the EP, an untriggered, subthreshold GRB was discovered in the Fermi/GBM data. The T$_{90}$, the duration containing $90\%$ of the burst's total photon counts, was measured as $\sim 50$ seconds in the band of $10-1000$ keV, and a bright flux was estimated at $4.5(+0.5,-0.5) \times 10^{-8}$ erg/s/cm$^2$\citep{2024GCN.37563....1L,2024GCN.37580....1R}. INTEGRAL SPI-ACS and SVOM/GRM also detected this event at energies above $80$keV and $30-300$ keV, respectively \citep{2024GCN.37574....1S,2024GCN.37573....1R}.

A radio observation was performed by the Australia Telescope Compact Array (ATCA) at multiple frequencies, which claims to have a $\sim73.5\mu$Jy radio counterpart detected at 9 GHz with a consistent position \citep{2024GCN.37952....1G}. Several optical and infrared follow-ups have been carried out, but no significant detection was reported\citep{2024GCN.37566....1K,2024GCN.37567....1M,2024GCN.37570....1L,2024GCN.37575....1A,2024GCN.37578....1Z,2024GCN.37579....1B,2024GCN.37583....1B,2024GCN.37610....1S,2024GCN.37952....1G}. Here we present a two-detector (HL) analysis for EP240919a due to the unavailability of Virgo at that time. A p-value of $0.62$ suggests that we are likely detecting the background, and an exclusion distance can be found in Table \ref{tab:exclusion distances}.

\subsubsection{EP250108a}

EP240108a is an interesting source that coincides with the optical transient AT2025kg (or SN2025kg)\citep{2025A&A...702L..18R,2025ApJ...988L..60S,2025arXiv250718544Z,2025ApJ...988L..14E,2025arXiv250417034L}. The first detection of this source was by the EP Wide-field X-ray Telescope on 2025-01-08 at 12:30:28.34, given the position of R.A., Dec.$=55.623, -22.509$ deg with $2.2$ arcmin uncertainty. The X-ray event lasted for over $2500$ seconds, with an unabsorbed flux ($0.5-4$keV) of $4.2 (+1.2/-0.9) \times 10^{-11}$ erg/s/cm$^2$ \citep{2025GCN.38861....1L}. A follow-up by the EP Follow-up X-Ray Telescope was performed $22.2$ hours after the trigger, which resulted in a non-detection \citep{2025GCN.38888....1L}.

The first optical counterpart of EP240108a was identified by the 2m Liverpool telescope \citep{2025GCN.38878....1E}. The team observed a transient within the EP error region and measured a g' band magnitude of $20.10\pm0.06$. Further optical photometric follow-ups illustrated the light curve of this source \citep{2025GCN.38885....1Z,2025GCN.38891....1X,2025GCN.38902....1M,2025GCN.38907....1K,2025GCN.38912....1I,2025GCN.38914....1Z,2025GCN.38972....1S,2025GCN.38983....1E,2025GCN.39002....1R,2025GCN.39020....1P}, while spectroscopic observations indicated that this source is a broad-lined SN Ic at a redshift of $0.17641$ \citep{2025GCN.38908....1Z,2025GCN.38984....1X,2025GCN.38987....1L}. Here, we report the X-pipeline analysis results for the event EP250108a, for which the associated p-value was estimated to be 0.763, a very likely background noise value. Given the redshift of $0.17641$, the corresponding luminosity distance, under the same cosmology assumed in this work (see section \ref{EP240414a}), is $880.6$ Mpc. As indicated in Table \ref{tab:exclusion distances}, the actual distance is $\sim2.6$ times farther than our limited sensitivity.

Because the observed distance to this source is also determined, we can similarly estimate the upper limit on $E_{\mathrm{GW,90\%}}$ for the CSG waveforms. The allowed $E_{\mathrm{GW,90\%}}$ and $h_{\mathrm{rss,90\%}}$ are summarized in Table\,\ref{tab:energy limits-250108}.

\begin{table}
    \centering
    \caption{The upper limits on $E_{\mathrm{GW,90\%}}$ and $h_{\mathrm{rss,90\%}}$ for the CSG and CBC waveforms for the search of EP250108a.}
    \label{tab:energy limits-250108}
    \begin{tabular}{lccc}
        \hline
        Waveform & $f_{\mathrm{central}}$ (Hz) & $E_{\mathrm{GW,90\%}}$ ($M_{\odot}c^{2}$) & $h_{\mathrm{rss,90\%}}$ (Hz$^{-1/2}$) \\
        \hline
        \multicolumn{4}{l}{\textbf{CSG}} \\
        CSG & 70  & $6.715\times10^{-2}$ & $4.574\times10^{-23}$ \\
        CSG & 100 & $1.169\times10^{-1}$ & $4.224\times10^{-23}$ \\
        CSG & 150 & $1.437\times10^{-1}$ & $3.122\times10^{-23}$ \\
        CSG & 300 & $5.506\times10^{-1}$ & $3.056\times10^{-23}$ \\
        \hline
        \multicolumn{4}{l}{\textbf{CBC}} \\
        BNS  & -- & -- & $1.132\times10^{-22}$ \\
        NSBH & -- & -- & $2.177\times10^{-22}$ \\
        \hline
    \end{tabular}
\end{table}

\subsubsection{EP250125a}

The event EP240125a started on 2025-01-25T02:36:19, observed by the EP Wide Field Telescope at R.A., Dec.$=175.364, -21.708$ deg with a $2.3$ arcmin error radius. The transient was observed for $74$ seconds and then discarded for the automatic follow-up observation by the Follow-up X-Ray Telescope. A better constrained position was then presented as R.A., Dec.$=175.3639, -21.7138$ deg with $20$ arcsec uncertainty. The unabsorbed flux observed by the Wide Field Telescope ($0.5-4$keV) and the Follow-up X-Ray Telescope ($0.5-10$ keV) was also measured as $1.8 (-0.5/+0.7) \times 10^{-9}$ and $5.4 (-0.6/+0.6) \times 10^{-12}$ erg/s/cm$^2$ respectively \citep{2025GCN.39028....1W}.

A team conducted photometric and spectroscopic observations of EP240125a using the Gemini South telescope $4.43$ hours after the EP trigger. They successfully observed a new point-like object in both r and z bands, which fits within the EP positional error. A spectrum taken with the B480 grism presents a strong, wide absorption around $4730\r{A}$, which was later suggested to be the damped Lyman-alpha system. Together with several other narrow band features, the redshift was derived as z$=2.89$\citep{2025GCN.39027....1L}. The REM $60$ cm robotic telescope also presented an r-band detection \citep{2025GCN.39031....1F}, while further optical follow-ups reported non-detections\citep{2025GCN.39092....1G,2025GCN.39032....1A}.

Additionally, X-ray follow-ups were also conducted. Swift/XRT performed observations $49$ minutes after the EP trigger and found a bright X-ray source lying $2.1$ arcsec away from the EP-reported position and $3.8$ arcsec away from the optical counterpart\citep{2025GCN.39029....1K} Interestingly, by comparing the two XRT observations around the first trigger and $8.9$ hours later, the peak flux dropped significantly from $4.3 \pm1.0 \times10^{-12}$ to $4.8\pm1.1\times 10^{-13}$ erg cm$^{-1}$ s$^{-1}$\citep{2025GCN.39029....1K}. On the other hand, Fermi-GBM and Swift/BAT did not find any possible counterparts \citep{2025GCN.39039....1R,2025GCN.39060....1R}. The X-pipeline result showed a p-value of $0.16$, which unfortunately indicated background noise. 
The exclusion distances are summarized in Table \ref{tab:exclusion distances}, where the largest value, $243.9$ Mpc, is obtained with the sgc70Q9 waveform. By considering the luminosity distance of the optical counterpart, a z $=2.89$ is equivalent to $24877.7$ Mpc under the same cosmological parameters as EP240414a (see section \ref{EP240414a}), which is completely unable to detect a GW for LVK in this distance.

Here, we establish the $E_{\mathrm{GW,90\%}}$ upper limits for the CSG waveforms based on the known host distance and $h_{\mathrm{rss,90\%}}$ as presented in Table\,\ref{tab:energy limits-250125}. The $h_{\mathrm{rss,90\%}}$ for CBC waveforms are also provided.

\begin{table}
    \centering
    \caption{The upper limits on $E_{\mathrm{GW,90\%}}$ and $h_{\mathrm{rss,90\%}}$ for the CSG and CBC waveforms for the search of EP250125a.}
    \label{tab:energy limits-250125}
    \begin{tabular}{lccc}
        \hline
        Waveform & $f_{\mathrm{central}}$ (Hz) & $E_{\mathrm{GW,90\%}}$ ($M_{\odot}c^{2}$) & $h_{\mathrm{rss,90\%}}$ (Hz$^{-1/2}$) \\
        \hline
        \multicolumn{4}{l}{\textbf{CSG}} \\
        CSG & 70  & $1.034\times10^{2}$ & $6.339\times10^{-23}$ \\
        CSG & 100 & $1.349\times10^{2}$ & $5.068\times10^{-23}$ \\
        CSG & 150 & $2.298\times10^{2}$ & $4.409\times10^{-23}$ \\
        CSG & 300 & $8.005\times10^{2}$ & $4.114\times10^{-23}$ \\
        \hline
        \multicolumn{4}{l}{\textbf{CBC}} \\
        BNS  & -- & -- & $1.599\times10^{-22}$ \\
        NSBH & -- & -- & $2.981\times10^{-22}$ \\
        \hline
    \end{tabular}
\end{table}

\subsection{Statistical Probability}

\begin{figure}
    \centering
    \includegraphics[width=0.9\linewidth]{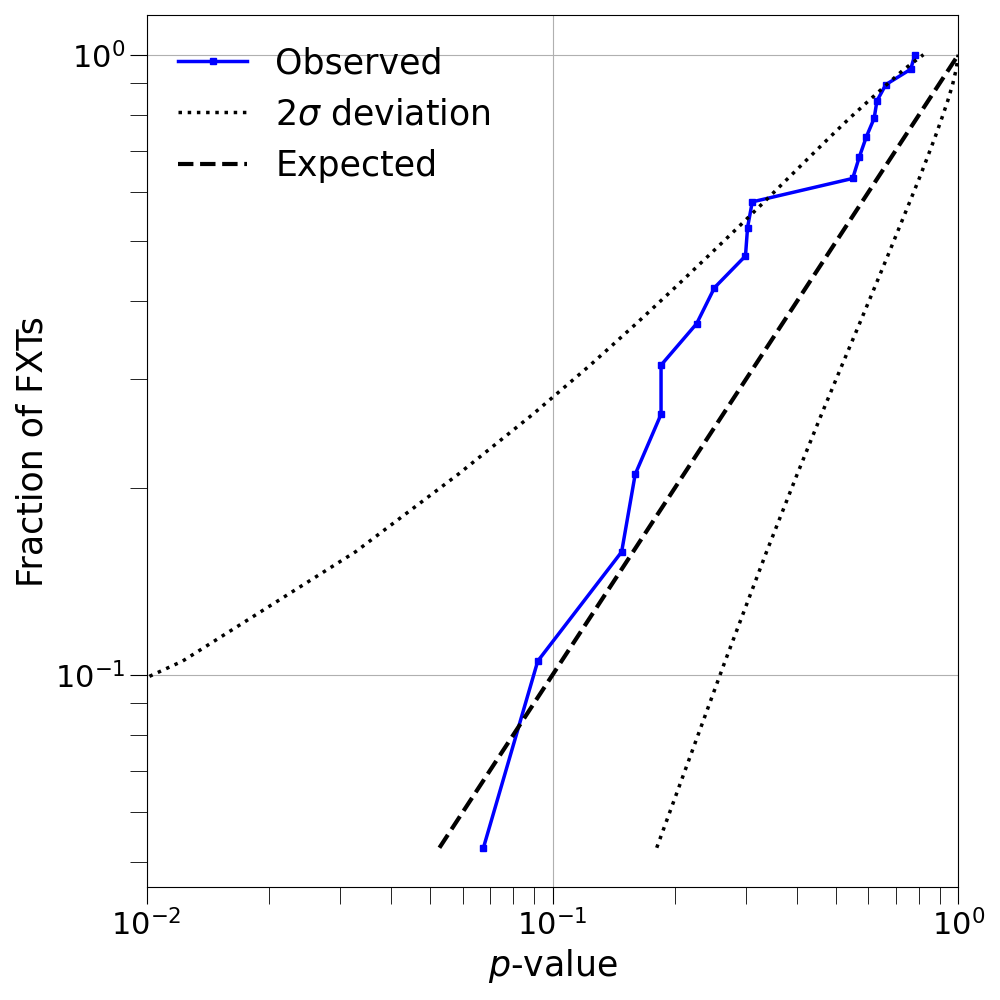}
    \caption{The p-value to FXT-fraction plot. The blue points presented the fraction of FXT as a function of p-value of all $19$ sources, and the dotted line showed the expected distribution of the null hypothesis, i.e., background.}
    \label{fig:p-value}
\end{figure}

\begin{figure}
    \centering
    \includegraphics[width=\linewidth]{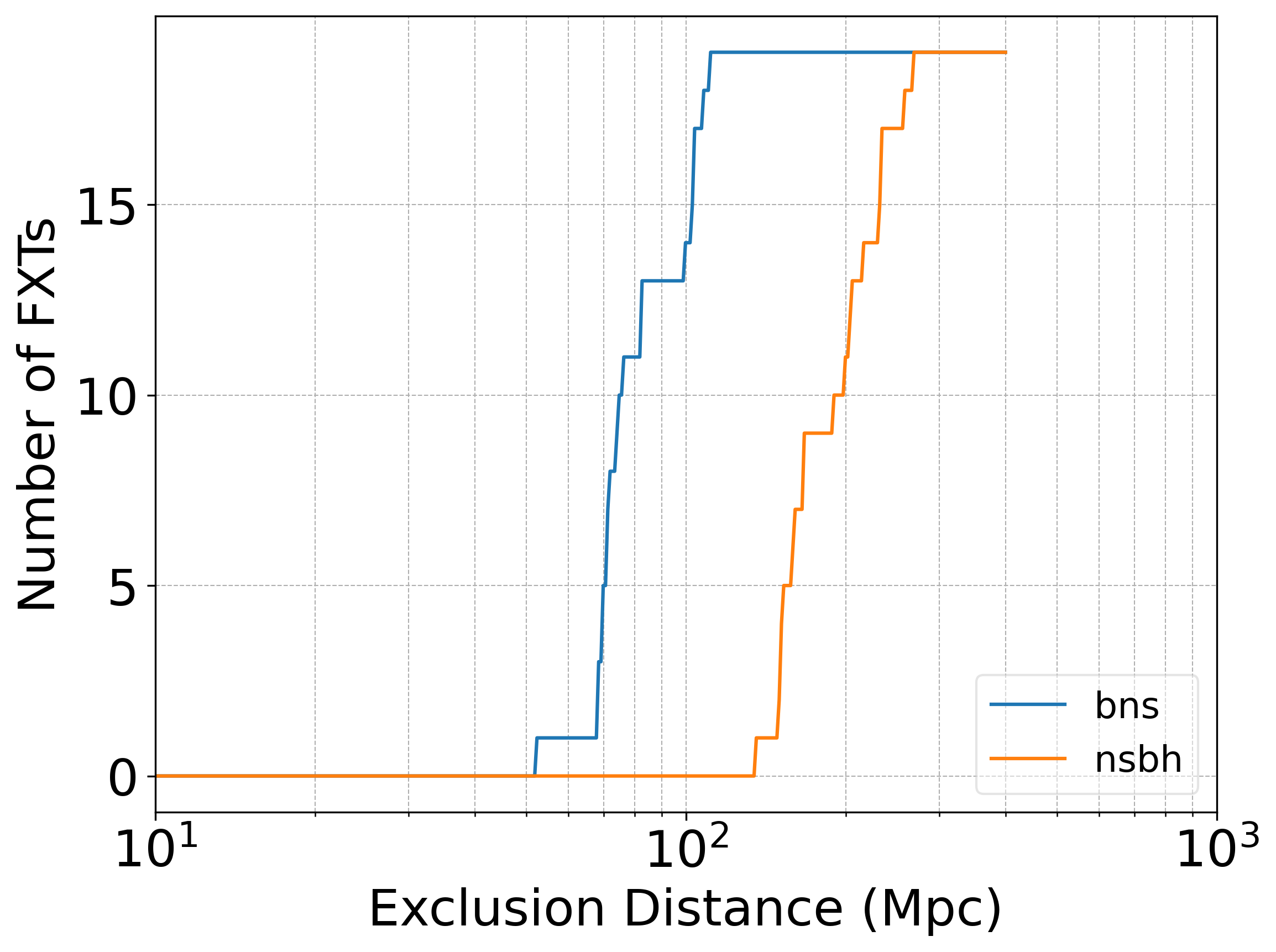}
    \caption{Cumulative histogram of the $90\%$ confidence exclusion distances for the $19$ FXTs that the X-pipeline searched.}
    \label{fig:exclusion distance}
\end{figure}

Figure~\ref{fig:p-value} presents the cumulative probability distributions of the X-pipeline results. The expected background is shown as the dashed line, while the $2\sigma$ error is represented by the dotted line. As all the estimated values were compatible with the background fluctuations, we indicated that no GW events were found in our samples. Furthermore, we illustrated the cumulative $90\%$ exclusion distance for all 19 FXTs, as presented in Figure~\ref{fig:exclusion distance}. Unlike what has been shown in the previous GRB search papers \citep{2021ApJ...915...86A, 2022ApJ...928..186A}, our result shows a greater difference between the Binary Neutron Star (BNS) merger and the Neutron Star-Black Hole (NSBH) scenario,  demonstrating a steady sensitivity of NSBH but slightly weaker sensitivity of BNS in our search.

\begin{table*}[htp]
    \centering
    \caption{p-values estimated under the null hypothesis for all $19$ sources.}
    \label{tab:p-value}
    \begin{tabular}{cccc}
        % \toprule
        \hline
        source name & p-value & interferometer network & redshift measurement\\
        \hline
        EP240413a & 0.227 & HLV & ...\\
        EP240414a & 0.068 & HLV & 0.401\\
        EP240426a & 0.662 & HLV & ...\\
        EP240426b & 0.570 & HL & ...\\
        EP240506a & 0.299 & HLV & ...\\
        EPJ0052.9-7230 & 0.148 & HLV&\\
        EP240617a & 0.593 & HLV & ...\\
        EP240625a & 0.781 & HLV & ...\\
        EP240626a & 0.302 & HLV & ...\\
        EP240708a & 0.251 & HLV & ...\\
        % EP240913a & 0.422 & HL & ...\\
        EP240918c & 0.185 & HL & ...\\
        EP240919a & 0.620 & HL & ...\\
        EP241125a & 0.185 & HL & ...\\
        EP241201a & 0.311 & HL & ...\\
        % EP241213a & 0.709 & HLV & ...\\
        EP241212a & 0.092 & HLV &...\\
        EP241231b & 0.630 & HL & ...\\
        EP250101a & 0.549 & HLV & ...\\
        EP250108a & 0.763 & HL & 0.176\\
        EP250125a & 0.160 & HL & 2.89\\
        \hline
    \end{tabular}
\end{table*}

\section{Conclusion} \label{sec:conclu}
In this work, we carried out a targeted search for GW signals from FXTs discovered by the Einstein Probe during the O4b run. During the LVK O4b operation, a total of $64$ FXTs were recorded, with $11$ occurring while all three GW detectors were operational and $8$ when both LIGO detectors were observing.
Using the unmodeled X-pipeline algorithm, the results show that none of the FXTs have significant GW signals. A $90$\% confidence level exclusion distance is set for each source according to the upper limits of possible waveform models, which are listed in Table \ref{tab:exclusion distances}.
Among the sources being analyzed, EP240414a is the most interesting one due to its identification as an Ic-BL supernova. While our measurement gives a $\sim200$ Mpc exclusion distance, optical observations show that it is located at a z$=0.41$ galaxy. This location is significantly beyond the sensitivity range that the LVK network could reach at this time.
It is worth noting that this work is based on the association between FXTs and GRBs. Further studies could be conducted to perform a more comprehensive search. For instance, we could extend the on-source search window by considering other potential GW mechanisms. Additionally, the time difference between the gamma-ray and X-ray triggers of a GRB might lead to a different on-source window. Furthermore, we could conduct a search using a single detector to expand the FXT sample or include KAGRA data to enhance sensitivity.
The work will be carried out in the future as the observation of EP continues to capture FXTs with a high resolution and detection rate.

\begin{acknowledgments}
We thank the anonymous referee for the insightful suggestions.
This work is supported by the National Science and Technology Council of Taiwan under the grants 114-2112-M-007-033-MY3 and 114-2811-M-007-077. The authors are grateful for the computational resources provided by the LIGO Laboratory and supported by National Science Foundation Grants PHY-0757058 and PHY-0823459.
\end{acknowledgments}

%\facilities{}

\software{X-pipeline\citep{2010NJPh...12e3034S}, Python, Matplotlib}
\clearpage

\appendix
\section{FXT Sample List} \label{A:sample list}

The full list with 64 FXTs is presented from Table \ref{tab:all_1} to Table \ref{tab:all_2}. An event will be marked as a GRB if it is temporally and spatially correlated with any GRB events detected by other GRB missions such as Fermi/GBM. The GRB name and the classification (long or short) of such events are also listed in Table \ref{tab:grb_all}. The column T$_\text{FXT}-$T$_\text{GRB}$ presents the duration between the two detections, while positive values indicate that GRBs were triggered earlier, and negative values show that FXTs occurred earlier.

\begin{table}[htp]
    \centering
    \caption{All FXTs observed by the Einstein Probe within LVK O4b run (2024-04-10 15:00 UTC to 2025-01-28 17:00 UTC), sorted in ascending order based on detection time.}
    \label{tab:all_1}
    \begin{threeparttable}
    \begin{adjustbox}{width=\textwidth,center}
        \begin{tabular}{ccccccc}
            \toprule
            Source Name & Triggered time & R.A. & Dec & Positional Error & Reference\\
            & (UTC time) & (deg) & (deg) & (deg)\\
            \hline
            EP240413a & 2024-04-13T14:39:37 & 228.794 & -18.800 & 0.333 & \cite{2024GCN.36086....1L}\\
            EP240414a & 2024-04-14T09:50:12 & 191.498 & -9.695 & 0.050 & \cite{2024GCN.36091....1L}\\
            EP240416a & 2024-04-16T02:42:13 & 203.150 & -13.612 & 0.050 & \cite{2024GCN.36138....1C}\\
            EP240417a & 2024-04-17T15:12:33 & 177.442 & -15.438 & 0.050 & \cite{2024GCN.36161....1H}\\
            EP240420a & 2024-04-20T12:04:28 & 228.713 & 14.796 & 0.050 & \cite{2024GCN.36194....1Z}\\
            EP240426a & 2024-04-26T04:23:21 & 121.857 & -29.461 & 0.003 & \cite{2024GCN.36313....1S}\\
            EPJ0052.9-7230&2024-05-27T08:41:28&13.215& -72.494&0.050&\cite{2024ATel16631....1Y}\\
            EP240426b & 2024-04-26T14:19:06 & 173.787 & -40.741 & 0.050 & \cite{2024GCN.36330....1P}\\
            EP240506a & 2024-05-06T05:01:39 & 213.978 & -16.715 & 0.050 & \cite{2024GCN.36405....1L}\\
            %EP240518a & 2014-05-14T13:02:08 & 216.955 & -49.565 & 0.050\\
            EP240617a\tnote{$^\dagger$} & 2024-06-17T12:19:13 & 285.030 & -22.561 & 0.050 & \cite{2024GCN.36691....1Z}\\
            EP240618a & 2024-06-18T05:43:43 & 281.627 & 23.820 & 0.050 & \cite{2024GCN.36690....1S}\\
            EP240625a & 2024-06-25T01:48:23 & 310.760 & -15.966 & 0.033 & \cite{2024GCN.36757....1P}\\
            EP240626a & 2024-06-26T06:28:28 & 263.023 & -13.051 & 0.033 & \cite{2024GCN.36766....1W}\\
            EP240702a & 2024-07-02T00:50:05 & 328.203 & -38.980 & 0.050 & \cite{2024GCN.36801....1C}\\
            EP240703a\tnote{$^\dagger$} & 2024-07-03T00:38:40 & 273.803 & -9.681 & 0.050 & \cite{2024GCN.36807....1W}\\
            EP240703b & 2024-07-03T05:24:26 & 279.539 & -57.401 & 0.050 & \cite{2024GCN.36810....1P}\\
            EP240703c & 2024-07-03T18:15:00 & 289.264 & -30.325 & 0.050 & \cite{2024GCN.36818....1Z}\\
            EP240708a & 2024-07-08T23:28:23 & 345.963 & -22.840 & 0.050 & \cite{2024GCN.36838....1C}\\
            EP240801a & 2024-08-01T09:06:03 & 345.140 & 32.610 & 0.040 & \cite{2024GCN.36997....1Z}\\
            EP240802a\tnote{$^\dagger$} & 2024-08-02T10:32:52 & 287.807 & -2.312 & 0.003 & \cite{2024GCN.37019....1W}\\
            EP240804a\tnote{$^\dagger$} & 2024-08-04T21:36:53 & 337.664 & -39.124 & 0.067 & \cite{2024GCN.37034....1W}\\
            EP240806a & 2024-08-06T04:47:53 & 11.491 & 5.091 & 0.035 & \cite{2024GCN.37063....1W}\\
            EP240807a\tnote{$^\dagger$} & 2024-08-07T05:42:48 & 300.970 & -68.777 & 0.042 & \cite{2024GCN.37088....1F}\\
            EP240809a&2024-08-09T16:53:08&268.259&-45.868&0.037&\cite{2024ATel16765....1L}\\
            EP240816b & 2024-08-16T01:44:27 & 16.013 & 15.398 & 0.056 & \cite{2024GCN.37185....1Y}\\
            EP240816a & 2024-08-16T03:28:42 & 292.925 & -54.412 & 0.044 & \cite{2024GCN.37188....1Y}\\
            EP240820a & 2024-08-20T00:54:47 & 16.246 & -34.694 & 0.035 & \cite{2024GCN.37214....1L}\\
            EP240908a & 2024-09-08T18:49:05 & 14.003 & 8.073 & 0.003 & \cite{2024GCN.37432....1M}\\
            EP240913a\tnote{$^\dagger$} & 2024-09-13T11:39:33 & 16.681 & 16.750 & 0.042 & \cite{2024GCN.37492....1L}\\
            EP240918a & 2024-09-18T11:24:37 & 289.394 & 46.128 & 0.006 & \cite{2024GCN.37541....1Z}\\
            EP240918b & 2024-09-18T15:40:00 & 258.660 & 66.739 & 0.048 & \cite{2024GCN.37555....1L}\\
            EP240918c & 2024-09-18T18:06:47 & 281.338 & -13.167 & 0.038 & \cite{2024GCN.37555....1L}\\
            \hline
        \end{tabular}
        \end{adjustbox}
        \begin{tablenotes}
            \footnotesize
            \item[$\dagger$] GRB correlated sources, detail listed in Table\ref{tab:grb_all}.
        \end{tablenotes}
    \end{threeparttable}
\end{table}

\begin{table}[htp]
    \centering
    \caption{(Continue from Table\ref{tab:all_1}) All FXTs observed by the Einstein Probe within LVK O4b run (2024-04-10 15:00 UTC to 2025-01-28 17:00 UTC), sorted in ascending order based on detection time.}
    \label{tab:all_2}
    \begin{adjustbox}{width=\textwidth,center}
    \begin{threeparttable}
        \begin{tabular}{ccccccc}
            \toprule
            Source Name & Triggered time & R.A. & Dec & Positional Error & Reference\\
            & (UTC time) & (deg) & (deg) & (deg)\\
            \hline
            EP240919a\tnote{$^\dagger$} & 2024-09-19T14:47:40 & 334.280 & -9.736 & 0.006 & \cite{2024GCN.37561....1L}\\
            EP240930a\tnote{$^\dagger$} & 2024-09-30T17:17:50 & 319.899 & 41.303 & 0.035 & \cite{2024GCN.37648....1T}\\
            EP241021a & 2024-10-21T05:07:56 & 28.852 & 5.957 & 0.040 & \cite{2024GCN.37834....1H}\\
            EP241025a\tnote{$^\dagger$} & 2024-10-25T01:36:45 & 333.641 & 83.577 & 0.003 & \cite{2024GCN.37864....1L}\\
            EP241026b & 2024-10-26T18:14:30 & 56.403 & 41.031 & 0.048 & \cite{2024GCN.37902....1L}\\
            EP241026a\tnote{$^\dagger$} & 2024-10-26T22:41:28 & 293.401 & 57.986 & 0.003 & \cite{2024GCN.37909....1L}\\
            EP241101a & 2024-11-01T23:52:49 & 37.763 & 22.731 & 0.047 & \cite{2024GCN.38039....1L}\\
            EP241103a & 2024-11-03T01:24:37 & 27.757 & 18.959 & 0.006 & \cite{2024GCN.38051....1Z}\\
            EP241104a\tnote{$^\dagger$} & 2024-11-04T18:30:15 & 32.574 & 31.555 & 0.045 & \cite{2024GCN.38081....1Z}\\
            EP241107a & 2024-11-07T14:10:23 & 35.009 & 3.333 &	0.003 & \cite{2024GCN.38112....1Z}\\
            % EP241109a & 2024-11-09T06:01:55 & 18.3599 & 0.0184 & 0.005\\
            EP241113b\tnote{$^\dagger$} & 2024-11-13T11:22:50 & 110.233 & 46.800 & 0.042 & \cite{2024GCN.38214....1L}\\
            EP241113a & 2024-11-13T19:12:53 & 131.996 & 52.382 & 0.003 & \cite{2024GCN.38211....1L}\\
            EP241115a\tnote{$^\dagger$} & 2024-11-15T05:47:20 & 19.416 & -17.954 & 0.043 & \cite{2024GCN.38239....1H}\\
            EP241119a & 2024-11-19T17:53:20 & 84.116 & 3.832 & 0.0383 & \cite{2024GCN.38281....1Z}\\ 
            EP241125a & 2024-11-25T00:06:06 & 48.561 & 37.677 & 0.044 & \cite{2024GCN.38318....1W}\\
            EP241126a & 2024-11-26T19:39:41 & 3.744 & 11.705 & 0.040 & \cite{2024GCN.38335....1H}\\
            EP241201a & 2024-12-01T21:01:22 & 282.596 & 66.081 & 0.039 & \cite{2024GCN.38415....1C}\\
            EP241202b & 2024-12-02T15:12:55 & 45.302 & 2.441 & 0.043 & \cite{2024GCN.38426....1Z}\\
            EP241206a & 2024-12-06T16:34:47 & 34.702 & 38.914 & 0.063 & \cite{2024GCN.38457....1Y}\\
            EP241208a & 2024-12-08T16:36:13 & 127.812 & 49.082 & 0.066 & \cite{2024GCN.38477....1W}\\
            EP241212a&2024-12-12T11:21:52&153.817&60.068&0.042&\cite{2024ATel16943....1L}\\
            EP241213a\tnote{$^\dagger$} & 2024-12-13T02:17:15 & 116.182 & 35.271 & 0.047 & \cite{2024GCN.38554....1Z}\\
            EP241217a & 2024-12-17T05:36:03 & 46.957 & 30.901 & 0.047 & \cite{2024GCN.38586....1Z}\\
            EP241217b\tnote{$^\dagger$} & 2024-12-17T16:58:04 & 84.167 & -25.281 & 0.047 & \cite{2024GCN.38606....1Z}\\
            EP241223a & 2024-12-23T07:21:23 & 74.804 & 7.11 & 0.055 & \cite{2024GCN.38660....1Z}\\
            EP241231b & 2024-12-31T19:27:29 & 100.064 & 16.171 & 0.043 & \cite{2025GCN.38778....1L}\\
            EP250101a&2025-01-01T05:52:21&85.575&0.352&0.053&\cite{2025GCN.38778....1L}\\
            EP250108a$^*$ & 2025-01-08T12:30:28 & 55.623 & -22.509 & 0.037 & \cite{2025GCN.38861....1L}\\
            EP250109a\tnote{$^\dagger$} & 2025-01-09T06:17:58 & 88.806 & -12.5 & 0.050 & \cite{2025GCN.38864....1L}\\
            EP250109b&2025-01-09T08:06:40&118.611&-14.651&0.055&\cite{2025ATel16974....1C}\\
            EP250111a & 2025-01-11T01:20:44 & 97.181 & 56.898 & 0.006 & \cite{2025GCN.38905....1Z}\\
            EP250125a & 2025-01-25T02:37:31 & 175.364 & -21.708 & 0.038 & \cite{2025GCN.39028....1W}\\
            \hline
        \end{tabular}
        \begin{tablenotes}
            \footnotesize
            \item[$\dagger$] GRB correlated sources, detail listed in Table\ref{tab:grb_all}.
        \end{tablenotes}
    \end{threeparttable}
    \end{adjustbox}
\end{table}

\begin{table}[htp]
    \centering
    \caption{All FXTs correlated to GRBs sorted ascending by detection time.}
    \label{tab:grb_all}
    \begin{adjustbox}{width=\textwidth,center}
    \begin{threeparttable}
        \begin{tabular}{cccccc}
            \toprule
            FXT Name & GRB Name & GRB triggered time & T$_\text{FXT}-$T$_\text{GRB}$ & Classification & Reference \\
                    &       &   (UTC time) & (second) & (detector) \\
            \hline
            EP240617a\tnote{*} & - & 2024-06-17T12:18:40 & 33
 & long(Fermi/GBM) & \cite{2024GCN.36692....1Y}\\
            EP240703a & GRB240703A & 2024-07-03T00:38:32 & 8 & long (Konus-Wind) & \cite{2025ApJ...993...51W}\\
            EP240802a & GRB240802A & 2024-08-02T10:34:03 & -71 & long (AstroSat CZTI) & \cite{2024GCN.37021....1A}\\
            EP240804a & GRB240804B & 2024-08-04T21:36:48 & 5 & long (Konus-Wind) & \cite{2024GCN.37071....1F}\\
            EP240807a & GRB240807A & 2024-08-07T05:42:45 & 3 & long (SVOM-GRM) & \cite{2024GCN.37097....1H}\\
            EP240913a & GRB240913C & 2024-09-13T11:42:33 & -180 & long (Fermi/GBM) & \cite{2024GCN.37493....1Y}\\
            EP240919a & GRB240919A & 2024-09-19T14:49:00 & -80 & long (Fermi/GBM) & \cite{2024GCN.37563....1L}\\
            EP240930a & GRB240930B & 2024-09-30T17:18:02 & -12 & long (Fermi/GBM) & \cite{2024GCN.37654....1D}\\
            EP241025a & GRB241025A & 2024-10-25T01:36:50 & -5 & long (Fermi/GBM) & \cite{2024GCN.37864....1L}\\
            EP241026a & GRB241026A & 2024-10-26T22:42:31 & -63 & long (Fermi/GBM) & \cite{2024GCN.37909....1L}\\
            EP241104a & GRB241104A & 2024-11-04T18:30:15 & 0 & long (Fermi/GBM) & \cite{2024GCN.38081....1Z}\\
            EP241113b & GRB241113B & 2024-11-13T11:23:05 & -15 & long (SVOM/ECLAIRs) & \cite{2024GCN.38214....1L}\\
            EP241115a & GRB241115D & 2024-11-15T05:45:36 & 104 & long (SVOM/ECLAIRs) & \cite{2024GCN.38250....1S}\\
            EP241213a & GRB241213A & 2024-12-13T02:19:00 & -105 & long (Swift/BAT) & \cite{2024GCN.38556....1B}\\
            EP241217b & GRB241217A & 2024-12-17T16:57:13 & 51 & long (SVOM/ECLAIRs) & \cite{2024GCN.38606....1Z}\\
            EP250109a & GRB250109A & 2025-01-09T06:17:08 & 49 & long (Fermi/GBM) & \cite{2025GCN.38873....1F}\\            
            \hline
        \end{tabular}
        \begin{tablenotes}
            \footnotesize
            \item[*] There was an untriggered detection reported afterward \citep{2024GCN.36692....1Y}.
        \end{tablenotes}
    \end{threeparttable}
    \end{adjustbox}
\end{table}

% \section{Rotating tables} \label{sec:rotate}

% To place a single page table in a landscape mode start the table portion with {\tt\string\begin\{rotatetable\}} and end with {\tt\string\end\{rotatetable\}}.bib

% Tables that exceed a print page take a slightly different environment since both rotation and long table printing are required. In these cases start with {\tt\string\begin\{longrotatetable\}} and end with {\tt\string\end\{longrotatetable\}}. The {\tt\string\movetabledown} command can be used to help center extremely wide, landscape tables. The command {\tt\string\movetabledown=1in} will move any rotated table down 1 inch. 

% A handy "cheat sheet" that provides the necessary \latex\ to produce 17
% different types of tables is available at \url{http://journals.aas.org/authors/aastex/aasguide.html#table_cheat_sheet}.

\clearpage
\bibliography{references}{}
\bibliographystyle{aasjournalv7}

\end{document}